# N-Graphdiyne two-dimensional nanomaterials: Semiconductors with low thermal conductivity and high stretchability

Bohayra Mortazavi[*,1], Meysam Makaremi[2], Masoud Shahrokhi[3], Zheyong Fan[4] and Timon Rabczuk[5, #]

[1]Institute of Structural Mechanics, Bauhaus-Universität Weimar, Marienstr. 15, D-99423 Weimar, Germany.
[2]Department of Materials Science and Engineering, University of Toronto, 184 College Street, Suite 140, Toronto, ON M5S 3E4, Canada.
[3]Institute of Chemical Research of Catalonia, ICIQ, The Barcelona Institute of Science and Technology, Av. Països Catalans 16, ES-43007 Tarragona, Spain.
[4]QTF Centre of Excellence, Department of Applied Physics, Aalto University, FI-00076 Aalto, Finland.
[5]College of Civil Engineering, Department of Geotechnical Engineering, Tongji University, Shanghai, China.

## Abstract

Most recently, N‐graphdiyne two-dimensional (2D) nanomaterials were successfully experimentally realized at the gas/liquid and liquid/liquid interfaces. We accordingly conducted density functional theory (DFT) and molecular dynamics simulations to explore the mechanical/failure, thermal conductivity and stability, electronic and optical properties of three N-graphdiyne nanomembranes. Our DFT results of uniaxial tensile simulations reveal that these monolayers can yield remarkably high stretchability or tensile strength depending on the atomic structure and loading direction. Studied N-graphdiyne nanomembranes were found to exhibit semiconducting electronic character, with band-gap values ranging from 0.98 eV to 3.33 eV, based on the HSE06 estimations. The first absorption peak suggests that these 2D structures can absorb visible, IR and NIR light. Ab initio molecular dynamics results reveal that N-graphdiyne 2D structures can withstand at high temperatures, like 2000 K. Thermal conductivities of suspended single-layer N-graphdiyne sheets were predicted to be almost temperature independent and about three orders of magnitude smaller than that of the graphene. The comprehensive insight provided by this work highlights the outstanding physics of N-graphdiyne 2D nanomaterials, and suggest them as highly promising candidates for the design of novel stretchable nanodevices.

Corresponding authors: *bohayra.mortazavi@gmail.com,
#timon.rabczuk@uni-weimar.de

## 1. Introduction

Since 2004 that the isolation of graphene from graphite was reported [1,2], the interest toward the two-dimensional (2D) materials has kept increasing. Graphene, the most prominent member of 2D nanomaterials exhibits outstanding physics, including interesting electronic and optical properties along with ultrahigh mechanical properties [3] and thermal conductivity [4]. These remarkable material properties propose the graphene as an exceptional candidate for the various applications, such as the mechanically robust and flexible components [5], heat management nanodevices [6,7], nanoelectronics [8], supercapacitors [9] and nanooptics [10]. Nevertheless, graphene exhibits zero band-gap semiconducting electronic character, and thus in its pristine form face limitations to be exploited as a 2D transistor [11]. This electronic nature of graphene however has been acting positively, in the sense that has promoted and oriented the researches toward the synthesis of novel 2D nanostructures with inherent semiconducting electronic properties, like transition metal dichalcogenides [12-14] and phosphorene [15,16].

Experimental realization of carbon based 2D materials with semiconducting electronic character has nevertheless attracted remarkable attentions during the last few years. In this regard, among various solutions, 2D nanomaterials with covalent networks of carbon and nitrogen atoms have been among the most successful paths to fabricate 2D semiconductors. For example, graphitic carbon nitride g-$C_3N_4$ [17,18] nanomembranes with semiconducting electronic character have been successfully synthesized, which have proven to illustrate highly desirable performances for various applications, such as the energy storage and conversion, fuel cells, catalysis, photocatalysis and $CO_2$ capture [17,19-24]. Nitrogenated holey graphene with ordered distributed holes and nitrogen atoms and a $C_2N$ stoichiometry, is another attractive member of carbon-nitride nanomaterials family, which was successfully synthesized via a simple wet-chemical reaction [25]. Recently, graphene like carbon-nitrogen semiconducting structure, 2D polyaniline crystals with $C_3N$ stoichiometry was experimentally realized via the direct pyrolysis of hexaaminobenzene trihydrochloride single crystals in solid state [26].

Apart from the carbon-nitride 2D semiconductors [25-29], graphyne [30] structures provide another class of full carbon allotropes, which are made from sp and sp$^2$ hybrid bonded atoms arranged in crystal lattices. In the work by Baughman *et. al* [30] published in 1987, numerous graphyne structures were proposed, some presenting semiconducting electronic nature. Interestingly, three decades after the theoretical perditions, two graphyne structures have been recently experimentally synthesized. In this regard, Jia *et al.* [31] in 2017 reported the fabrication of carbon Ene-yne graphyne from tetraethynylethene by solvent-phase reaction. Shortly after, the synthesize of crystalline graphdiyne nanosheets were reported by Matsuoka *et al.* [32]. Graphyne structures have been widely theoretically explored and they are predicted to yield highly attractive properties, suitable for diverse applications, such as; anode material for metal-ion batteries [33,34], hydrogen storage [35-38], catalysts [39], thermoelectricity [40,41] and nanotransistors [42-45].

Most recently, an exciting experimental advance has just taken place with respect to the synthesis of N-graphdiyne nanomembranes at the gas/liquid and liquid/liquid interfaces [46]. N-graphdiyne 2D sheets are likely to the graphyne structures; however, the nitrogen atoms replace 3 sp carbon atoms in the connecting hexagonal rings. This latest experimental success in the fabrication of N-graphdiyne [46] nanomembranes highlights the importance of theoretical studies in order to provide understanding of their intrinsic material properties. Such that comprehensive analysis of structural, thermal, mechanical, optical and electronic properties of 2D N-graphdiyne structures plays critical roles in the design of advanced nanodevices exploiting the outstanding properties of these novel nanosheets. The objective of the present investigation is therefore to efficiently explore the properties of three different novel N-graphdiyne structures through extensive atomistic simulations. To this aim, first-principles density functional theory (DFT) simulations were employed to investigate mechanical, thermal stability, optical and electronic properties of these newly synthesized nanomembranes. The phononic thermal conductivity of these systems was also studied using the equilibrium classical molecular dynamics simulations. This work provides a general vision concerning the critical properties of a new class of 2D semiconductors and such that we hope that the acquired results can guide future theoretical and experimental studies.

## 2. Computational methods

Density functional theory simulations were carried out by using the Vienna *Ab initio* Simulation Package (VASP) [47-49] including the Perdew-Burke-Ernzerhof (PBE) functional [50] for the exchange correlation potential. The interaction between the valence and core electrons was described on the basis of the projected augmented wave (PAW) method [51]. A plane-wave cutoff energy of 500 eV was used for the valence electrons. The VESTA [52] package was employed for visualization of the optimized structures. Fig.1, illustrates unit-cells of three N-graphdiyne lattices as they were realized by Kan *et al.* [46]. In order to distinguish different structures, we name them according to their unit-cell atomic composition; $C_{18}N_6$, $C_{12}N_2$, and $C_{36}N_6$. In order to analyze the anisotropicity in the transport properties, we studied the properties along the armchair and zigzag directions, as depicted in Fig. 1. These directions are defined in analogy to graphene, on the basis of hexagonal rings connecting the carbon chains. For the evaluation of optical responses, the $x$ and $y$ directions (as shown in Fig. 1) were considered according to the unit-cell vectors. In all simulations, periodic boundary conditions were applied along all three Cartesian directions and a vacuum thickness of 16 Å was considered to avoid image-image interactions along the sheet normal direction.

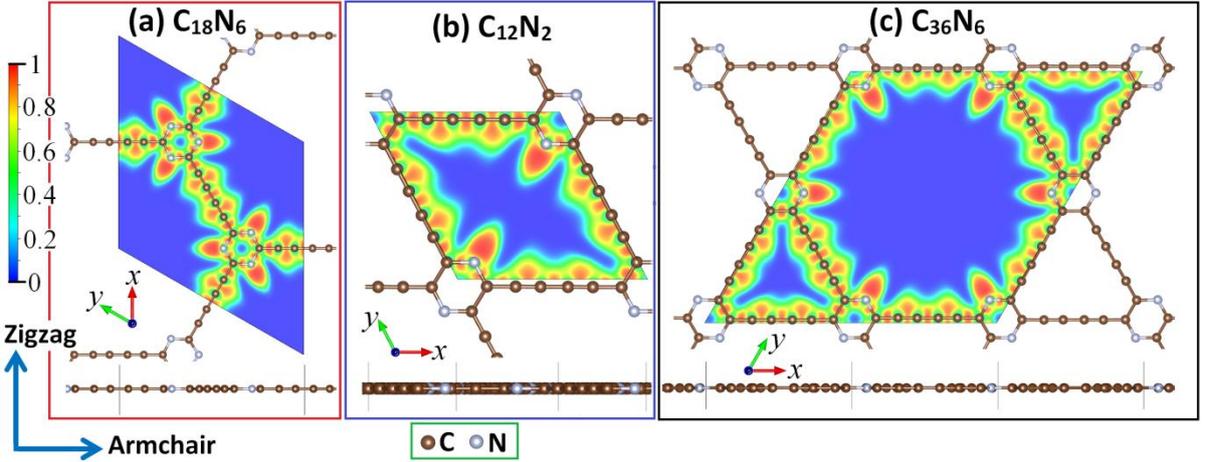

Fig. 1, Top and side views of the atomic structure of $C_{18}N_6$, $C_{12}N_2$, and $C_{36}N_6$ N-graphdiyne monolayers studied in this work. Contours illustrate electron localization function within the unit-cells. Mechanical and thermal conduction properties were studied along the armchair and zigzag directions as shown. Contours illustrate electron localization function (ELF), which takes a value between 0 and 1, where 1 corresponds to the perfect localization and ELF = 0.5 corresponds to the electron gas.

In order to evaluate the mechanical properties, we conducted uniaxial tensile simulations. In this case, only unit-cell structures were considered for $C_{18}N_6$ and $C_{36}N_6$ as they are shown in Fig. 1 and for the single-layer $C_{12}N_2$ we however simulated a rectangular unit-cell. We explored the anisotropy in the mechanical responses by conducting the uniaxial tensile simulations along the armchair and zigzag directions. For each case, the periodic simulation box size along the loading direction was increased gradually with a engineering strain step of 0.001. In order to satisfy the uniaxial tensile loading condition, the stress values along the transverse directions of loading should be negligible during the tensile deformation. Since the atoms are in contact with vacuum along the thickness, the stress is negligible in this direction. Therefore, to satisfy the uniaxial tensile loading condition, the simulation box size along the width was changed with a goal to reach negligible stresses (below 0.04 N/m). To avoid any sudden bond stretching or void formation, the atomic positions were rescaled after applying the changes in the simulation box size. To simulate the atomic rearrangements during the tensile simulations, the conjugate gradient method was employed for the geometry optimizations, with termination criteria of $10^{-4}$ eV and 0.01 eV/Å for the energy and the forces, respectively, using a 5×5×1 Monkhorst-Pack [53] k-point mesh size. To evaluate the thermal stability, ab initio molecular dynamics (AIMD) simulations were carried out using the Langevin thermostat with a time step of 1fs and 2×2×1 k-point mesh size.

Since the PBE functional underestimates the band gap values, we also employed the screened hybrid functional, HSE06 [54] to evaluate the electronic properties of these monolayers, with Γ-centred Monkhorst-Pack mesh size of 8×8×1. We used 14×14×1 k-point grids for computing the optical properties. Optical properties, including the imaginary and real parts of dielectric and absorption coefficient were calculated through the random phase approximation (RPA) method [55] constructed over the PBE results. Optical properties were described by photon frequency dependent dielectric function, $\varepsilon(\omega) = \mathrm{Re}\,\varepsilon_{\alpha\beta}(\omega) + i\,\mathrm{Im}\,\varepsilon_{\alpha\beta}(\omega)$, which is mainly acquired from the electronic structures. The imaginary part of the dielectric function for semiconductors could be obtained by only taking into account the contribution of interband transition contribution [56,57]:

$$\mathrm{Im}\,\varepsilon_{\alpha\beta}(\omega) = \frac{4\pi^2 e^2}{\Omega} \lim_{q \to 0} \frac{1}{|q|^2} \sum_{c,v,k} 2w_k \delta(\varepsilon_{ck} - \varepsilon_{vk} - \omega) \times \left\langle u_{ck+e_\alpha q}|u_{\alpha k}\right\rangle \left\langle u_{ck+e_\beta q}|u_{\alpha k}\right\rangle^* \quad (1)$$

In this equation, $q$ is the Bloch vector of the incident wave and $w_k$ is the k-point weight. The band indices $c$ and $\alpha$ are restricted to the conduction and the valence band states, respectively. The vectors $e_\alpha$ are the unit vectors for the three Cartesian directions and $\Omega$ is the volume of the unit-cell. $u_{ck}$ is the cell periodic part of the orbitals at the $k$-point k. The real part $\mathrm{Re}\,\varepsilon_{\alpha\beta}(\omega)$ can be evaluated from $\mathrm{Im}\,\varepsilon_{\alpha\beta}(\omega)$ using the Kramers–Kronig transformation [56]:

$$\mathrm{Re}\,\varepsilon_{\alpha\beta}(\omega) = 1 + \frac{2}{\pi} P \int_0^\infty \frac{\omega'\,\mathrm{Im}\,\varepsilon_{\alpha\beta}(\omega')}{(\omega')^2 - \omega^2 + i\eta}\,d\omega' \qquad (2)$$

where $P$ denotes the principle value and $\eta$ is the complex shift in Kramers-Kronig transformation. The absorption coefficient was calculated using the following relation [58]:

$$a_{\alpha\beta}(\omega) = \frac{2\omega k_{\alpha\beta}(\omega)}{c} \qquad (3)$$

where $k_{\alpha\beta}$ is imaginary part of the complex refractive index and $c$ is the speed of light in vacuum, known as the extinction index. $k_{\alpha\beta}$ was acquired according to:

$$k_{\alpha\beta}(\omega) = \sqrt{\frac{\left|\varepsilon_{\alpha\beta}(\omega) - \mathrm{Re}\,\varepsilon_{\alpha\beta}(\omega)\right|}{2}} \qquad (4)$$

The optical spectra of N-graphdiyne nanosheets have been obtained for the in-plane and out-of-plane directions in order to assess the anisotropicity of optical properties.

We employed classical molecular dynamics (MD) simulations to evaluate the lattice thermal conduction properties of N-graphdiyne nanosheets at different temperatures. We used an efficient code, GPUMD (*Graphics Processing Units Molecular Dynamics*) [59] to compute the lattice thermal conductivity on the basis of equilibrium molecular dynamics (EMD) method. In this work the atomic interactions were introduced by employing the Tersoff potential [60] with the optimized parameters by Lindsay and Broido [61] and Kinarci *et al.* [62] for carbon-carbon and carbon-nitrogen interactions, respectively. Worthy to note that despite of the fact that the employed potential functions in this work are theoretically sound for the modelling of heat transfer in carbon based 2D materials and they have been used in numerous most recent molecular dynamics studies [63-68], the investigation of impact of potential function choice on the predicted thermal conductivity of N-graphdiyne nanomembranes is an important topic for the future studies. In the EMD method, the heat flux vector was calculated with the appropriate form for many-body potentials [69,70], via:

$$\boldsymbol{J} = \sum_i \sum_{j \neq i} (\boldsymbol{r}_j - \boldsymbol{r}_i)\left(\frac{\partial U_j}{\partial \boldsymbol{r}_{ji}} \cdot \boldsymbol{v}_i\right), \qquad (5)$$

where $\boldsymbol{v}_i$ is the velocity of atom $i$, $\boldsymbol{r}_i$ is the position vector of atom $i$ and $U_j$ is the potential energy associated with atom $j$. From heat-flux, the lattice thermal conductivity tensor can be obtained from the Green-Kubo formula:

$$\kappa_{\alpha\beta} = \frac{1}{V k_B T^2} \int_0^\infty \langle \boldsymbol{J}_\alpha(0)\boldsymbol{J}_\beta(t)\rangle\,dt, \qquad (6)$$

where $k_B$ is Boltzmann's constant, $T$ is the temperature, and $V$ is the total volume of the monolayer, assuming a thickness of 3.35 Å. In our EMD simulations, we constructed relatively large samples with more than 20000 atoms. We first equilibrate the system in the NVT ensemble with the target temperature for 250 ps and then sample the heat flux in the NVE ensemble for 250 ps. The velocity-Verlet integration method [71,72] with a time step of 0.25 fs used in all simulations. For each sample at a given temperature, we conducted five independent simulations with uncorrelated initial velocities to improve the statistical accuracy and calculate an error for a time-converged thermal conductivity as the standard error of the values obtained from the five independent runs.

## 3. Results and discussions

In the Fig. 1, top and side views of the energy minimized N-graphdiyne monolayers are illustrated. The hexagonal lattice constants for the $C_{18}N_6$, $C_{12}N_2$ and $C_{36}N_6$ were found to be 16.038 Å, 7.985 Å and 18.664 Å, respectively. For the $C_{12}N_2$ with rectangular unit-cell, the lattice lengths along the armchair and zigzag directions were calculated to be 9.67 Å and 15.97 Å, respectively. Interestingly, the corresponding bond lengths in different structures were found to be very close. In this regard, the C-C and C-N bond lengths in the connecting hexagonal rings were measured to be 1.44 Å and ~1.35 Å, respectively. For the carbon chains, the longest and smallest bond lengths were found to be 1.424 Å and 1.224 Å, respectively. In this case, the $sp^2$-sp carbon bonds are longer, but yet smaller than the $sp^2$-$sp^2$ bonds in the hexagonal rings. The atomic structures of energy minimized N-graphdiyne monolayers are given in detail in the supporting information of this manuscript. In Fig. 1, we also plotted the electron localization function (ELF) [73]. The ELF is a position-dependent function ranging from 0 to 1. In this case, the values of ELF close to one corresponds to the region with high probability of finding electron localizations. The ELF = 0.0 illustrates that the electrons are completely delocalized and these exist no electron. The ELF = 0.5 corresponds to the region of electron gas-like behaviour [73]. For the all of studied monolayers, the ELF values around the center of all C-C and C-N bonds are greater than 0.8, confirming the covalent bonding in these structures. Nevertheless, the electron localization is also considerable around the nitrogen atoms originated from their higher valance electrons as well as their higher electronegativity leading to charge gain from carbon atoms.

To examine the energetic stability of N-graphdiyne monolayers, the cohesive energies per atoms were calculated as defined by; $E_{coh} = -(\sum_i E_i - E_t)/n$, where $E_t$, $E_i$ and $n$ are the total energy of unit-cell, the energy of the $i\text{-}th$ isolated atom and the total number of atoms in the unit-cell, respectively. The cohesive energies for the $C_{18}N_6$, $C_{24}N_4$ and $C_{36}N_6$ monolayers were accordingly calculated to be -6.02 eV, -6.31 eV and -6.08 eV, respectively. The negative cohesive energies confirm the energetic stability of these monolayers in the free-standing state.

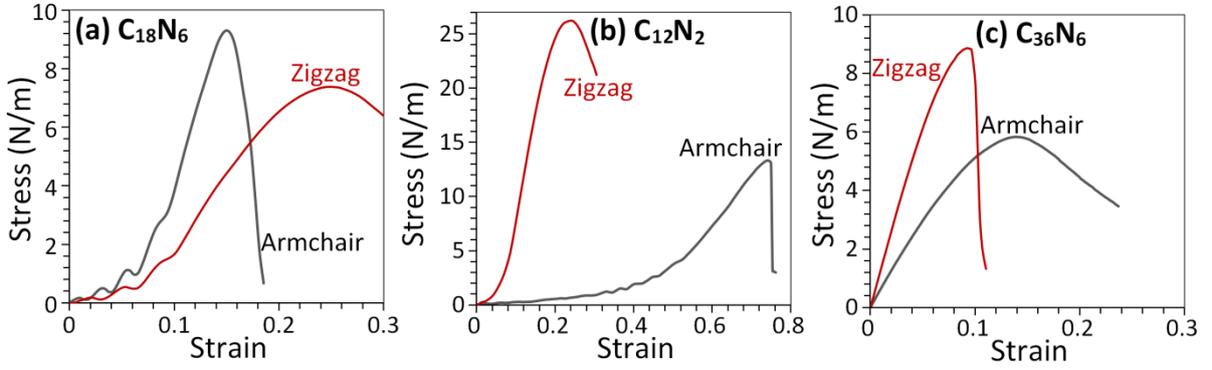

Fig. 2, Uniaxial stress-strain responses of single-layer N-graphdiyne structures along the armchair and zigzag directions.

We first study the mechanical properties of these novel carbon-nitride 2D nanostructures. In Fig. 2, the first-principles results for the uniaxial stress-strain responses of single-layer $C_{18}N_6$, $C_{12}N_2$, and $C_{36}N_6$ elongated along the armchair and zigzag directions are illustrated. For the most of materials with dense packing structures, the stress-strain curve shows an initial linear response corresponding to the linear elasticity. Such an initial relation however cannot be observed in the stress-strain curves for the $C_{18}N_6$ and $C_{12}N_2$ monolayers and thus they do not exhibit linear elasticity. The elastic response in the mechanical properties of packed structures materials is associated with the bond stretching. For example, for the pristine graphene, the stretching of the structure can be achieved only by increasing the carbon atoms bond length and such that at small strain levels the stress values increase linearly [74]. The unusual initial patterns in the stress-strain response of $C_{18}N_6$ and $C_{12}N_2$ can be attributed to the fact that at initial strain levels the stretching of these monolayers does not directly leads to the increase of bonds lengths. In these cases, the deformations is mainly achieved by the severe contraction of the structures along the perpendicular direction of loading (sheet's width) and since the bond elongations are limited the stress-values increase very slightly. Unlike the aforementioned two nanomembranes, the $C_{36}N_6$ monolayer presents elasticity which suggests that from the early stages of the loading, the deformation is achieved by the bond stretching rather than the structural deflection. In this monolayer, the triangular shaped chains resist against the sheet contraction along the perpendicular direction of loading and thus stiffen the structure.

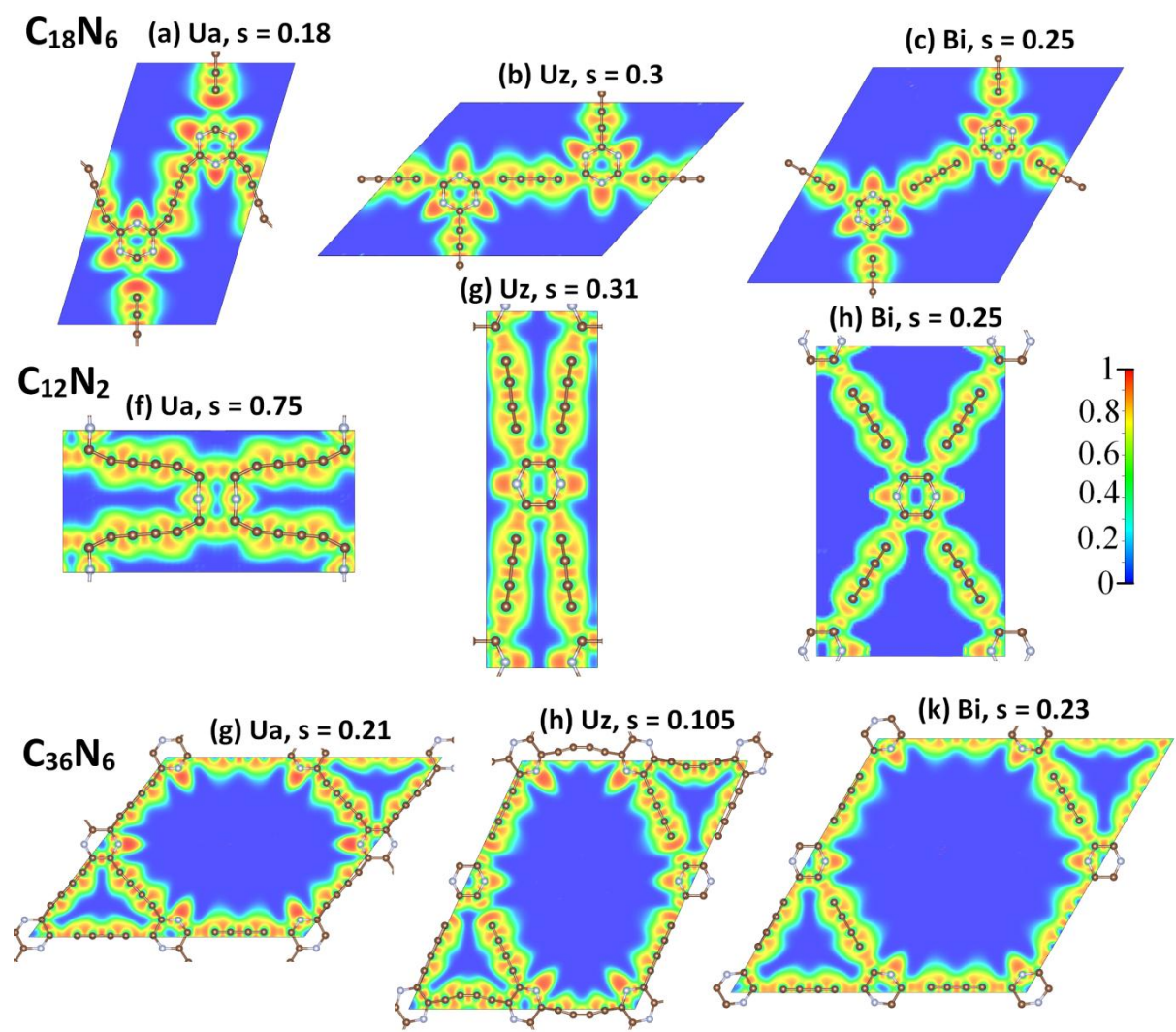

Fig. 3, Snapshots of deformed N-graphdiyne nanomembranes after the tensile strength point at different strain values. Here, s, Ua, Uz and Bi stand for the strain level, uniaxial loading along the armchair, uniaxial loading along the zigzag and biaxial loading, respectively. The length criteria for the bond illustrations was chosen to be 2 Å. Contours illustrate electron localization function (ELF). The ELF close to one occurring around the center of bonds confirm the intactness of the original covalent bonds, whereas the vanish of the ELF concentration between the connecting atoms and its consequent split suggests the destroying of the corresponding covalent bond.

In the case of $C_{18}N_6$ monolayer, along the armchair direction the structure yields higher stiffness as compared with the zigzag direction (see Fig. 2a). This can be explained because of the fact that for the uniaxial loading along the armchair, from every two carbon-carbon chains existing in the system, one is exactly oriented along the loading direction and thus the bonds are more involved in the load transfer. For this structure, stress fluctuations are observable for the strain levels below ~0.1 which suggest that the deformation process is not uniform. In these cases, as the loading proceeds the bond elongations are observable by the increase of the stress values, however as the contraction along the sheet's width occurs, it relives some parts of the stress in the bonds. After strain levels of ~0.1, the deformation of $C_{18}N_6$ monolayer is dominated by the bonds elongation and thus the stress values uniformly increase up to the ultimate tensile point, where the structure reaches its maximum load bearing limit. The ultimate tensile strength of $C_{18}N_6$ along the armchair and zigzag directions were predicted to be 9.3 N/m and 7.4 N/m, respectively. According to our calculations, the strain at the ultimate tensile strength was found to be 0.15 and 0.25 for the uniaxial loading along the armchair and zigzag directions, respectively. In Fig. 3a and Fig. 3b, snapshots of the deformed $C_{18}N_6$ are shown. As it is depicted in Fig. 3b, the carbon chains oblique to the loading direction with an angle of 30 degree, during the uniaxial loading along the zigzag direction rotate and finally become exactly in-line with the loading direction.

Among the studied N-graphdiyne nanosheets, $C_{12}N_2$ monolayer exhibits the highest anisotropicity in the mechanical properties. In this system, for the uniaxial loading along the zigzag direction the deflection along the sheet's width is highly limited and after the strain levels of ~0.05 the deformation is mostly achieved by the bond elongation. This monolayer along the zigzag direction exhibits a remarkably high tensile strength of 26.2 N/m, which is almost twice of that along the armchair direction, 13.3 N/m. Along the armchair direction the $C_{12}N_2$ monolayer illustrates an ultrahigh stretchability, keeping its load bearing ability up to strain levels of ~0.75, more than threefold larger as compared with the zigzag direction. The outstanding stretchability of $C_{12}N_2$ along the armchair direction is due to the fact that the carbon chains can more freely rotate, by almost by 60 degree (see Fig. 3f) as the structure approaches the ultimate tensile strength point. As shown in Fig. 3g, for the uniaxial loading along the zigzag direction the maximum rotation of carbon chains is less than 30 degree. The tensile strengths of $C_{12}N_2$ monolayer are distinctly higher than those of the $C_{18}N_6$ and $C_{36}N_6$. The carbon chains in the $C_{12}N_2$ monolayer illustrate the highest flexibility and by increasing the strain levels they all finally orient along the loading direction and involve directly in the load bearing. The more bonds involved in the lead transfer is equivalent with the higher rigidity and tensile strength as well. As depicted in Fig. 3f and Fig. 3g, the $C_{12}N_2$ structures can reach the highest packing densities under the uniaxial loading conditions.

Likely to other studied sheets, $C_{36}N_6$ nanosheet also exhibits anisotropic mechanical properties. Unlike the $C_{18}N_6$ structure, the $C_{36}N_6$ monolayer is found to be stiffer along the zigzag direction as compared with the armchair direction. In this structure, for the uniaxial loading along the armchair, from every three carbon chains in the system, only one chain is involved in the stretching, whereas for the uniaxial loading along the zigzag two chains are involve in the load bearing. According to our results, the elastic modulus of $C_{36}N_6$ monolayer along the zigzag and armchair directions was found to be 126 N/m and 63 N/m, respectively. As expected the elastic moduli of $C_{36}N_6$ monolayer are considerably smaller than that of the  graphene, predicted to be 350.7 N/m by Liu *et al.* [74]. The tensile strength of $C_{36}N_6$ was also estimated to be 5.8 N/m and 8.9 N/m, along the armchair and zigzag directions, respectively. Among the studied N-graphdiyne nanosheets, $C_{36}N_6$ structure exhibits the lowest stretchability. In $C_{36}N_6$ monolayer the stiffening effects of triangular shaped carbon chains limit the deflection of the structure. This limitation is more considerable for the loading along the zigzag, in which the carbon chains are obliquely oriented along the loading and can only slightly rotate to align with the loading direction. In this case the rest of carbon chains are exactly along the transverse direction of loading, and they limited ability for contraction is easily observable since they undergo buckling during the deformation process (see Fig. 3h). For the uniaxial loading of $C_{36}N_6$ along the armchair direction, the rotation of the carbon chains are easier and thus the structure can stretch up to higher strain levels (see Fig. 3g).

In order to briefly analyze the failure mechanism, snapshots of the deformed $C_{18}N_6$, $C_{12}N_2$, and $C_{36}N_6$ monolayers after their tensile strength points are compared in Fig. 3. In these cases, we also conducted the biaxial loading simulations to compare the results with those of the uniaxial. According to the results shown in Fig. 3, for the all studied samples the bond breakage happen in the $sp^2$-sp carbon bonds, with an exception for the $C_{12}N_2$ monolayer elongated along the armchair (Fig. 3f) in which the bond breakages occur for the $sp^2$-$sp^2$ carbon bonds in the hexagonal ring. For the all studied N-graphdiyne nanosheets under the biaxial loading, all $sp^2$-sp carbon bonds fail simultaneously (Fig. 3c, h and k). However, for the samples under the uniaxial loading condition, the bond failures happen for those oriented along the loading direction, which were involved in the stretching and load bearing as well. In these cases, no bond breakage is observable in the carbon chains oriented along the transverse direction of loading.

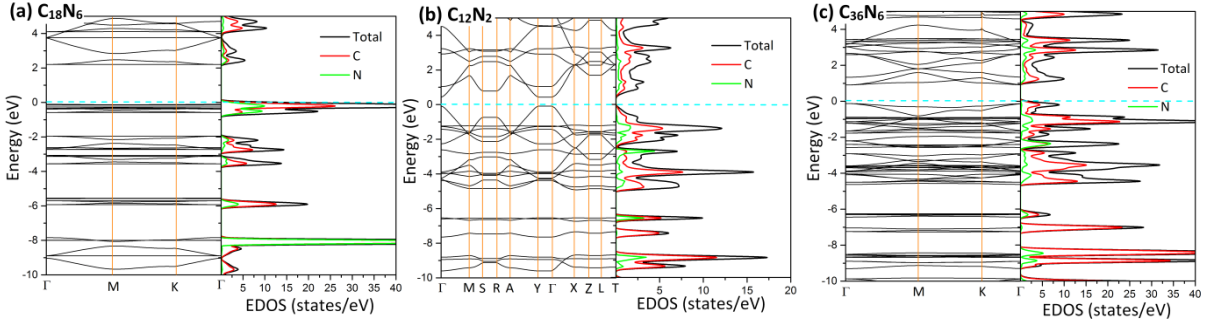

Fig. 4, Band structure, total and partial EDOS of free-standing $C_{18}N_6$, $C_{12}N_2$ and $C_{36}N_6$ monolayers predicted by the PBE functional. The Fermi energy is aligned to zero.

In order to evaluate the electronic properties, the band structure along the high symmetry directions and the total electronic density of states (EDOS) were calculated using the PBE functional. Fig. 4 illustrates the band structure, total and partial EDOS of stress-free $C_{18}N_6$, $C_{12}N_2$ and $C_{36}N_6$ monolayers. Interestingly, $C_{18}N_6$ and $C_{36}N_6$ monolayers exhibit direct band-gap which the valence band maximum (VBM) and the conduction band minimum (CBM) locate at $\Gamma$-point. According to the PBE results the band-gaps of the single-layer $C_{18}N_6$ and $C_{36}N_6$, are 2.20 eV and 1.10 eV, respectively. For the $C_{12}N_2$ monolayer the VBM and CBM are parallel and coincide at $Y$-$\Gamma$ direction with the band-gap of 0.5 eV. From partial EDOS it is visible that the valance band near to the Fermi level in all structures is dominated by density states of carbon atoms. The corresponding band-gap values within the HSE06 functional for unstrained $C_{18}N_6$, $C_{12}N_2$ and $C_{36}N_6$ monolayers are found to be 3.33 eV, 0.98 eV and 1.55 eV, respectively. To probe the possibility of the tuning of the electronic band-gap, the EDOS for these monolayers were also calculated under different magnitudes of biaxial and uniaxial loading strains using the PBE and the HSE06 hybrid functional and the obtained results are summarized in Table 1. In Fig. S1 the EDOS of stress-free and strained N-graphdiyne monolayers predicted by the HSE06 functional are illustrated. It was found that by applying the biaxial loading conditions the electronic band-gap of studied monolayers widens. Applying the uniaxial loading along the armchair and zigzag directions lead to decrease in the electronic band-gaps of $C_{18}N_6$ and $C_{36}N_6$ single-layers whereas they widen the band-gap of $C_{12}N_2$ monolayer. These results confirm the strain tuneable electronic character of N-graphdiyne nanomembranes. Among the studied structures, the single-layer $C_{12}N_2$ yields the narrowest band-gap, close to that of the silicon and such that can be potentially very attractive for the application in the post-silicon flexible nanoelectronics.

Table 1. Band-gap (eV) values of single-layer $C_{18}N_6$, $C_{12}N_2$ and $C_{36}N_6$ predicted by PBE and HSE06 methods.

| Band gap | $C_{18}N_6$ | | $C_{12}N_2$ | | $C_{36}N_6$ | |
|---|---|---|---|---|---|---|
| | PBE | HSE06 | PBE | HSE06 | PBE | HSE06 |
| Stress-free | 2.20 | 3.33 | 0.50 | 0.98 | 1.10 | 1.55 |
| Biaxial strain=0.05 | 2.36 | 3.43 | 0.95 | 1.50 | 1.29 | 1.92 |
| Biaxial strain=0.10 | 2.50 | 3.60 | 1.40 | 2.25 | 1.69 | 2.52 |
| Uniaxial-armchair, strain= 0.05 | 2.15 | 3.25 | 0.50 | 1.01 | 0.90 | 1.51 |
| Uniaxial-armchair, strain= 0.10 | 1.90 | 3.15 | 0.52 | 1.06 | 0.87 | 1.42 |
| Uniaxial-zigzag, strain= 0.05 | 2.20 | 3.27 | 0.50 | 1.13 | 0.89 | 1.30 |
| Uniaxial-zigzag, strain= 0.10 | 2.10 | 3.24 | 0.76 | 1.37 | 0.73 | 1.28 |

We next analyze the optical response of these novel 2D materials. The imaginary and real parts of the dielectric function of free-standing aforementioned systems for in-plane (($E//x$ and $E//y$) and out-plane directions ($E//z$), as it is shown in Fig. 1, obtained from PBE+RPA are given in Fig. 5. Since the optical spectra of $C_{18}N_6$ and $C_{36}N_6$ systems are isotropic for the light polarizations along the x- and y-axis, we just report the optical properties of these compounds for light polarizations along the x-axis. The absorption edge of $\mathrm{Im}\varepsilon_{\alpha\beta}(\omega)$ occurs at 2.15 eV, 0.20 eV and 0.75 eV for $C_{18}N_6$, $C_{12}N_2$ and $C_{36}N_6$ systems for $E//x$, respectively, which is in visible range for $C_{18}N_6$ monolayer. The corresponding values for $E//z$ are 7.91 eV, 6.12 eV and 6.42 eV. The absorption edge of $\mathrm{Im}\varepsilon_{\alpha\beta}(\omega)$ for $C_{12}N_2$ along y-axis is 0.85 eV. It was found that the static dielectric constant (the real part of dielectric constant at zero energy, Re $\varepsilon_0$) along $E//x$ is 2.03 eV, 13.72 eV and 4.77 eV for $C_{18}N_6$, $C_{12}N_2$ and $C_{36}N_6$ monolayers, respectively, and the corresponding values for $E//z$ are 1.09 eV, 1.21 eV and 1.07 eV. The Re $\varepsilon_0$ for $C_{12}N_2$ along $E//y$ is 10.07 eV which is smaller than that along $E//x$. It is known that the roots of Re $\varepsilon_0$ with x = 0 line shows the plasma frequencies [75]. The first plasma frequency for single-layer $C_{18}N_6$, $C_{12}N_2$ and $C_{36}N_6$ along $E//x$ was accordingly predicted to be 2.70 eV, 0.55 eV and 1.33 eV, respectively. The first plasma frequency for $C_{12}N_2$ monolayer along E||y is 1.34 eV.

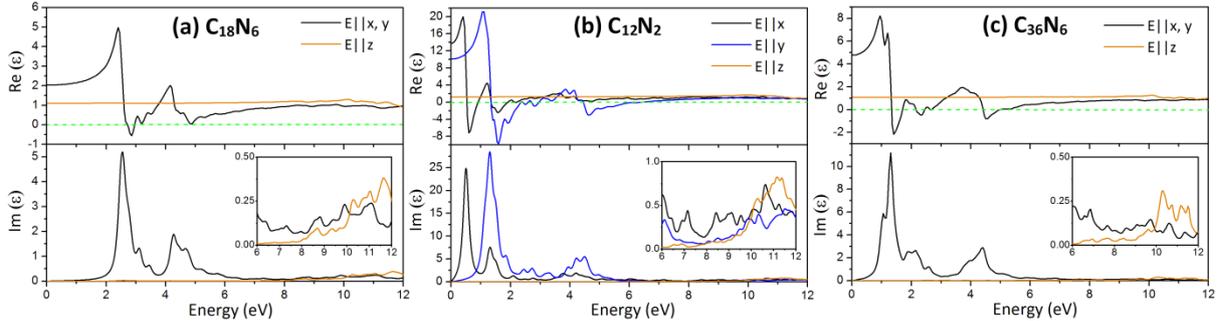

Fig. 5, Imaginary and real parts of the dielectric function of free-standing $C_{18}N_6$, $C_{12}N_2$ and $C_{36}N_6$ monolayers for the parallel ($E//x$ and $E//y$) and perpendicular ($E//z$) light polarizations, calculated using the PBE plus RPA approach.

The absorption coefficient $\alpha_{ij}(\omega)$ for all polarizations are plotted in fig. 6. In order to compare the optical properties of N-graphdiyne monolayers with graphene [76], absorption spectrum of pristine graphene for in-plane direction was also computed and depicted in Fig. 6 (with red dashed line). Our results show that the first absorption peak for $C_{18}N_6$ monolayer occurs at energy of 2.57 eV along $E//x$ which is desirable for the practical applications in optoelectronic devices in the visible spectral range. The first absorption peak of $C_{36}N_6$ monolayer for in-plane polarization locates at 1.1 eV which is in near IR (NIR) range of light. The corresponding values in $C_{12}N_2$ monolayer for $E//x$ and $E//y$ are 0.53 eV (in IR range) and 1.31 eV (in near IR range). The corresponding values of $C_{18}N_6$, $C_{12}N_2$ and $C_{36}N_6$ monolayers for perpendicular polarization are 8.10 eV, 6.0 eV and 6.93 eV, respectively, which are in ultraviolet range. The first absorption peak for graphene was found to locate at the energy of 0.72 eV, which is in good agreement with the previous work [77]. It is clear that the absorption coefficient of N-Graphdiyne structures for in-plane polarization is larger than that of the graphene, whilst for high frequency regime (greater than 12 eV) the absorption coefficient of graphene is larger than N-Graphdiyne structures. These results confirm that the optical absorption of N-Graphdiyne structures in the visible range is better than that of the graphene, which is promising for the applications in optoelectronics and nanoelectronics. In general, the results of optical properties for the studied nanomembranes confirm that the optical spectra are highly anisotropic along the in-plane and out-of-plane directions. Interestingly, anisotropic optical spectra along the in-plane ($E//x$ and $E//y$) directions in $C_{12}N_2$ can be promising for the design of novel electronic and optical nanodevices that exploit in-plane anisotropic optical properties, such as polarization-sensitive photodetectors [78]. The polarization sensitivity is due to the strong intrinsic linear dichroism, which arises from the in-plane optical anisotropy of these novel 2D nanomaterials [79]. Worthy to note that likely to other 2D materials, chemical functionalization with foreign atoms can be also employed to efficiently modify the electronic and optical properties of N-graphdiyne nanosheets, which is an attractive topic for the future studies.

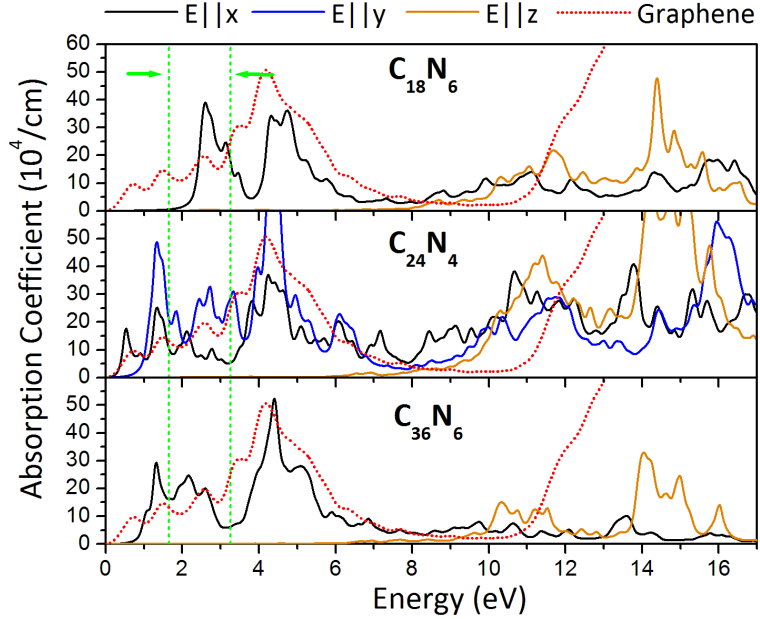

Fig. 6, Absorption coefficient of $C_{18}N_6$, $C_{12}N_2$ and $C_{36}N_6$ monolayers for the in-plane ($E//x$ and $E//y$) and out-of-plane ($E//z$) light polarizations. The red dashed line shows the in-plane absorption spectra for pristine graphene. The visible range of the spectrum is showed by green dashed lines.

The range of temperatures that a material can keep its structure intact is another important property that plays a critical role for the service at high temperatures. We accordingly explore the thermal stability of single-layer N-graphdiyne structures by performing the AIMD simulations. To this aim, AIMD simulations were conducted at different temperatures of 500 K, 1000 K, 1500 K, 2000 K and 2500 K for 10 ps. The snapshots of $C_{18}N_6$, $C_{12}N_2$, and $C_{36}N_6$ monolayers after the AIMD simulations are illustrated in Fig. 7. According to our results, all studied N-graphdiyne monolayers could endure at the high temperature of 2000 K. All the studied systems however face the bond breakages at the temperature of 2500 K. In this temperature, the $C_{12}N_2$, and $C_{36}N_6$ monolayers show a very similar failure mechanism, in which the rupture initiates in the sp$^2$-sp$^2$ carbon bonds. This observation is also consistent with our previous study on the carbon Ene-yne (CEY) graphyne [80], in which the first bond breakage at high temperatures was found to occur in the sp$^2$-sp$^2$ carbon bonds. As illustrated in Fig. 7c, in the case of $C_{18}N_6$ monolayer the first bond breakage occurred for the carbon-nitrogen bond. Our AIMD results reveal the outstanding thermal stability of N-graphdiyne monolayers and thus confirm their suitability for high temperature applications in nanotechnology [81-83]. Despite of remarkably high thermal stabilities and mechanical tensile strengths of N-graphdiyne monolayers, they may show dynamical instability, which suggest the analysis of phonon dispersions of these structures as an interesting topic for the future studies.

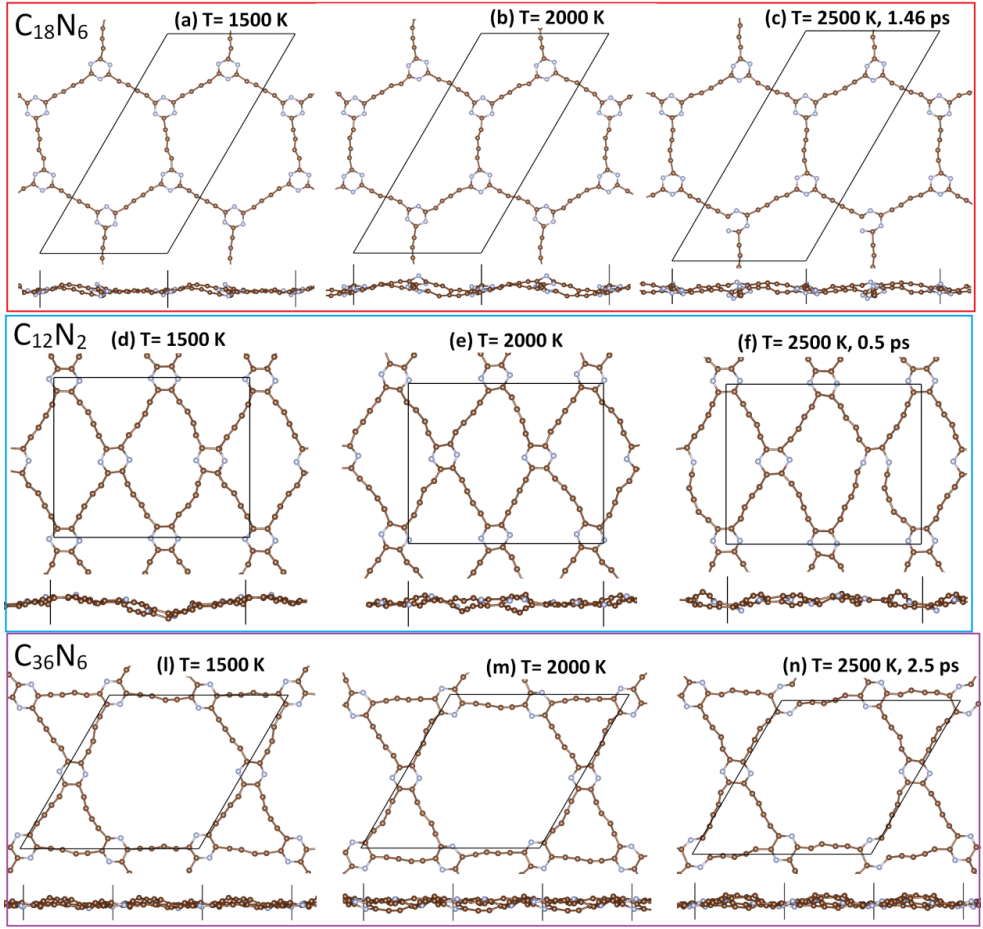

Fig. 7, Top and side snapshots of single-layer $C_{18}N_6$, $C_{12}N_2$ and $C_{36}N_6$ structures at different temperatures of 1500 K, 2000 K and 2500 K after the AIMD simulations for 10 ps.

Thermal conductivity is another important property of 2D materials in the design of nanodevices [6,7,84,85]. High thermal conductivity is favourable to avoid overheating issues, while, low thermal conductivity is desirable to improve figure of merit of thermoelectric materials. We accordingly also calculate the thermal conductivity of single-layer $C_{18}N_6$, $C_{12}N_2$, and $C_{36}N_6$ using the EMD method. In Fig. 8, the EMD predictions for the thermal conductivity of $C_{18}N_6$, $C_{12}N_2$, and $C_{36}N_6$ monolayers along the armchair and zigzag directions as a function of correlation time are illustrated.

As it can be observed, for all the studied monolayers the thermal conductivity well converges within about 1 ps. In contrast, the lattice thermal conductivity of pristine graphene, takes about 1 ns to converge to 2900±100 W/mK [70]. The results shown in Fig. 8 reveal that the thermal conductivities of single-layer $C_{18}N_6$, $C_{12}N_2$, and $C_{36}N_6$ are almost three orders of magnitude smaller than that of the pristine graphene [4,7,82,86,87]. It is also interesting to note that heat conduction is convincingly isotropic in $C_{18}N_6$ and $C_{36}N_6$ nanomemebranes but anisotropic in $C_{12}N_2$. At the room temperature, the thermal conductivities of $C_{18}N_6$ and $C_{36}N_6$ were predicted to be 1.36±0.06 W/mK and 2.39±0.12 W/mK, respectively. In the case of $C_{12}N_2$ monolayer the thermal conductivity along the zigzag direction is by around threefold higher. At the room temperature, the thermal conductivities of single-layer $C_{12}N_2$ along the zigzag and armchair directions were calculated to be 5.75±0.25 W/mK and 1.77±0.08 W/mK, respectively. The higher thermal conductivity along the zigzag direction can be simply understood by considering the carbon chains length that phonons have to travel along the heat transfer direction.

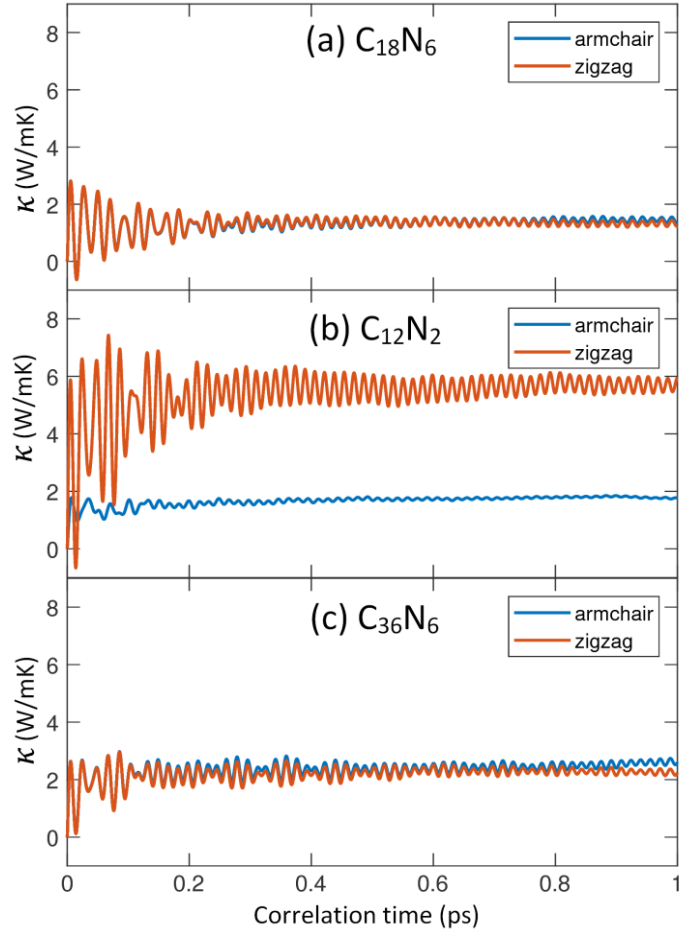

Fig. 8, Calculated thermal conductivity of single-layer $C_{18}N_6$, $C_{12}N_2$ and $C_{36}N_6$ at the room temperature along the armchair and zigzag directions as a function of correlation time.

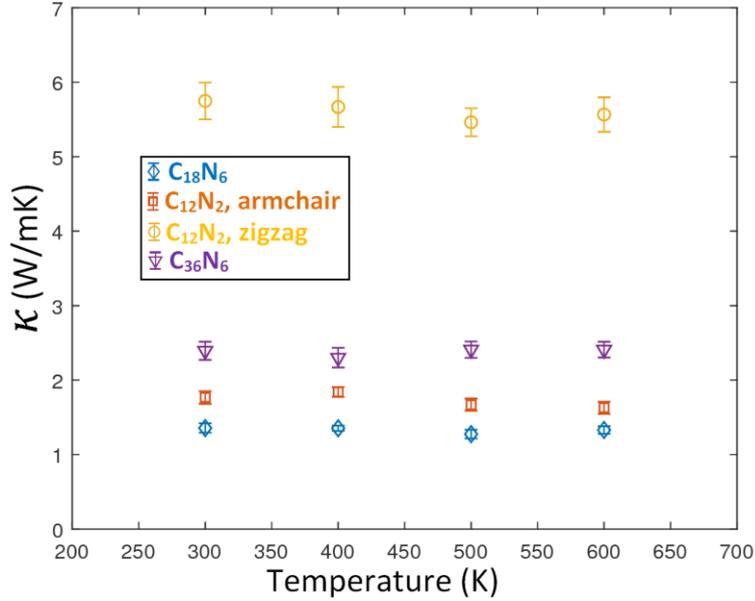

Fig. 9, EMD predictions for the thermal conductivity of single-layer $C_{18}N_6$, $C_{12}N_2$ and $C_{36}N_6$ at different temperatures.

In Fig. 9 the effect of temperature on the thermal conductivity of $C_{18}N_6$, $C_{12}N_2$, and $C_{36}N_6$ monolayers is demonstrated. Interestingly, our EMD results suggest that the thermal conduction in the N-graphdiyne nanomembranes are convincingly insensitive to the temperature. This is in sharp contrast with pristine materials with densely packed structures, where the thermal conductivity is inversely proportional to the temperature, due to the phonon-phonon scattering [88]. In the N-graphdiyne monolayers however the independency of thermal conductivity to the temperature implies that the contribution of phonon-phonon scattering is negligible in these systems [89]. For the thermal transport along defect-free graphene, out-of-plane phonons contribution is substantially higher than in-plane phonons [70,85]. According to the EMD results, for the pristine graphene the out-of-plane and in-plane thermal conductivities were estimated to be ~70% and ~30% of effective thermal conductivity, respectively [70]. In order to provide more insight concerning the heat transfer mechanism along the N-graphdiyne 2D structures, in Fig. 10 the out-of-plane and in-plane thermal conductivity components for single-layer $C_{18}N_6$ are compared. Notably, in these novel 2D nanomaterials the in-plane phonon dynamics dominate the thermal transport and the contribution of out-of-plane phonons is completely negligible. This observation suggests that the in-plane thermal conductivities we predicted for the single-layer N-graphdiynes are theoretically extendable to those of their multi-layers structures. The ultralow thermal conductivity along with the semiconducting electronic character may suggest the N-graphdiyne monolayer as promising candidates for the design of novel thermoelectric devices operating at low temperatures [90].

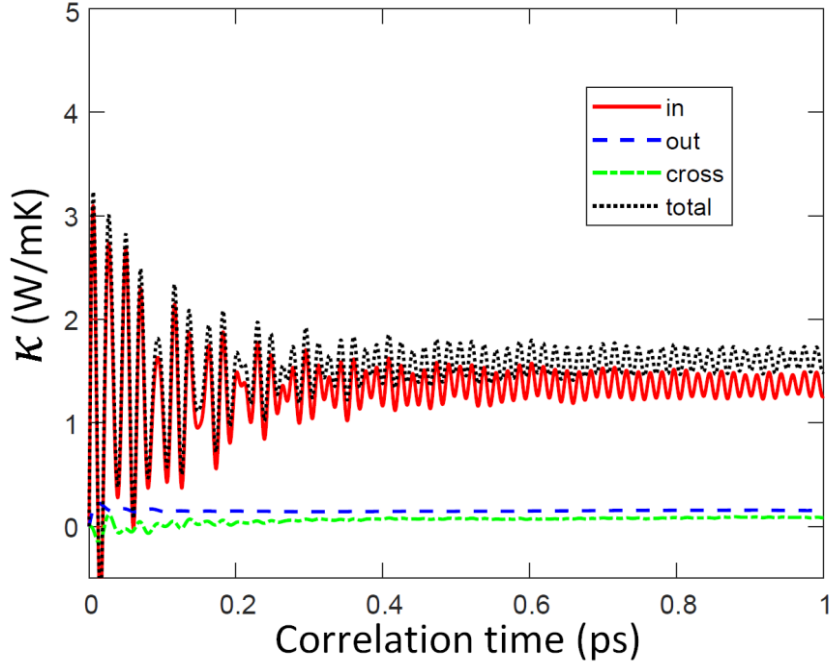

Fig. 10, EMD results for the out-of-plane (out), cross-plane (cross) and in-plane (in) thermal conductivity components for the single-layer $C_{18}N_6$ at the room temperature.

## 4. Conclusion

Most recently, novel 2D N-graphdiyne structures have been successfully experimentally realized at the gas/liquid and liquid/liquid interfaces. N-graphdiyne structures are expected to exhibit outstanding physics, originated from their low-density lattices made from carbon and nitrogen atoms. Motivated by recent experimental advances, we conducted extensive first-principles DFT and classical molecular dynamics simulations to explore the mechanical properties, thermal conductivity and stability, electronic and optical responses of free-standing and single-layer $C_{18}N_6$, $C_{12}N_2$, and $C_{36}N_6$ N-graphdiyne. Mechanical properties were evaluated by conducting the uniaxial tensile simulations using the DFT method. It was found that the N-graphdiyne monolayers exhibit highly anisotropic mechanical properties. The structural effects on the orientation dependent mechanical properties were discussed. Our first-principles results highlight that depending on the atomic structure and loading directions the N-graphdiyne monolayers can yield remarkably high stretchability or tensile strength. In the most of cases it was observed that the mechanical failure in the N-graphdiyne 2D structures initiates by the breakage of the

sp$^2$-sp carbon bonds. PBE and HSE06 functional calculations were employed to study the electronic properties of these novel nanomembranes. Our first-principles simulations confirm the semiconducting electronic character of all of the N-graphdiyne monolayers. The band-gap values of $C_{18}N_6$, $C_{12}N_2$ and $C_{36}N_6$ were predicted to be 3.33 eV, 0.98 eV and 1.55 eV, respectively, on the basis of HSE06 results. It was found that the application of biaxial strains widens the band-gap of all studied monolayers, whereas applying the uniaxial loading along the armchair and zigzag directions decrease the electronic band-gap of $C_{18}N_6$ and $C_{36}N_6$ and increase the band-gap of $C_{12}N_2$ monolayer. The dielectric tensor was derived within the random phase approximation. The first absorption peak reveal that these novel 2D nanostructures can absorb the visible, IR and NIR light, suggesting their prospect for the applications in optoelectronics and nanoelectronics. Ab initio molecular dynamics simulations confirm the outstanding thermal stability of N-graphdiyne monolayers, which can endure at high temperatures up to 2000 K. According to our classical equilibrium molecular dynamics simulations results, the thermal conductivities of single-layer $C_{18}N_6$, $C_{12}N_2$ and $C_{36}N_6$ at the room temperature were found to be three orders of magnitude smaller than that of the graphene, with maximum and minimum values of 5.75 W/mK and ~1.35 W/mK, respectively. It was shown that the thermal transport in these novel nanomembranes is convincingly insensitive to the temperature. The ultralow thermal conductivity along with the semiconducting electronic character may suggest the N-graphdiyne monolayers as promising candidates for the design of novel carbon based thermoelectric devices operating at low temperatures.

## Acknowledgment

B. M. and T. R. greatly acknowledge the financial support by European Research Council COMBAT project (Grant number 615132). Z. F. acknowledges the support from the Academy of Finland Centre of Excellence program (project 312298).

**Supporting Information**

**N-Graphdiyne two-dimensional materials: Semiconductors with low thermal conductivity and high stretchability**

Bohayra Mortazavi[*,1], Meysam Makaremi[2], Masoud Shahrokhi[3], Zheyong Fan[4] and Timon Rabczuk[5]

[1]*Institute of Structural Mechanics, Bauhaus-Universität Weimar, Marienstr. 15, D-99423 Weimar, Germany.*

[2]*Department of Materials Science and Engineering, University of Toronto, 184 College Street, Suite 140, Toronto, ON M5S 3E4, Canada.*

[3]*Institute of Chemical Research of Catalonia, ICIQ, The Barcelona Institute of Science and Technology, Av. Països Catalans 16, ES-43007 Tarragona, Spain.*

[4]*QTF Centre of Excellence, Department of Applied Physics, Aalto University, FI-00076 Aalto, Finland.*

[5]*College of Civil Engineering, Department of Geotechnical Engineering, Tongji University, Shanghai, China.*

*E-mail: bohayra.mortazavi@gmail.com

1. Atomic structures of N-Graphdiyne unit-cells in VASP POSCAR

2. HSE06 results for the EDOS of stress-free and strained N-Graphdiyne monolayers.

# 1. Atomic structures of N-Graphdiyne unit-cells in VASP POSCAR

## 1.1 C$_{18}$N$_6$

```
C18N6
   1.00000000000000
     16.0375168077986103    0.0000000000000000    0.0000000000000000
      8.0187584038993034   13.8888969726305600    0.0000000000000000
      0.0000000000000000    0.0000000000000000   16.0000000000000000
   C    N
   18    6
Direct
  0.3804121860736274  0.3804121974542980  0.6250000000000000
  0.4316795762031731  0.4316795914129870  0.6250000000000000
  0.4757496464837061  0.4757496360660340  0.6250000000000000
  0.5242503541734678  0.5242503770415965  0.6250000000000000
  0.5683204244540008  0.5683204216946365  0.6250000000000000
  0.6195877851269032  0.6195878156533254  0.6250000000000000
  0.2391756153880706  0.3804121974542980  0.6250000000000000
  0.3804121899027848  0.2391756250017565  0.6250000000000000
  0.6195878141557571  0.7608243813031450  0.6250000000000000
  0.7608243499211298  0.6195878156533254  0.6250000000000000
  0.5683204201970540  0.8633591692205371  0.6250000000000000
  0.5242503755440140  0.9514992585266029  0.6250000000000000
  0.4316795872628418  0.1366408302816495  0.6250000000000000
  0.4757496319158818  0.0485007409755767  0.6250000000000000
  0.1366429261230877  0.4316795846102650  0.6250000000000000
  0.0485049320680702  0.4757496292633121  0.6250000000000000
  0.8633570459888276  0.5683204148919145  0.6250000000000000
  0.9514950400438522  0.5242503702388746  0.6250000000000000
  0.2833474513407666  0.4333051021258214  0.6250000000000000
  0.4333050907451508  0.2833474588922726  0.6250000000000000
  0.2833474492785868  0.2833474588922726  0.6250000000000000
  0.5666948838567194  0.7166525474126288  0.6250000000000000
  0.7166525493164073  0.5666949109818020  0.6250000000000000
  0.7166525194319675  0.7166525474126288  0.6250000000000000
```

## 1.2 C$_{12}$N$_2$

```
C12N2
   1.00000000000000
     -4.8351798149264500    7.9850022016561102    0.0000000000000000
     -4.8351798149264500   -7.9850023845883298    0.0000000000000000
      0.0000000000000000    0.0000000000000000   16.0000000000000000
   C    N
   12    2
Direct
  0.9854693074224699  0.8082038266340206  0.6250000000000000
  0.9864911353807173  0.6579099403893380  0.6250000000000000
  0.9874938457420512  0.5266598865176277  0.6250000000000000
  0.9883546680330682  0.3832058352757386  0.6250000000000000
  0.9893573815515495  0.2519557374305705  0.6250000000000000
  0.9903792142847934  0.1016618913088365  0.6250000000000000
  0.1346532834451679  0.9573878221484691  0.6250000000000000
  0.2849471489507494  0.9563659700313707  0.6250000000000000
  0.4161971941103175  0.9553632479739989  0.6250000000000000
  0.5596512265774010  0.9545024036414489  0.6250000000000000
  0.6909013463226898  0.9534997090391926  0.6250000000000000
```

```
0.8411951941542881   0.9524778745960703   0.6250000000000000
0.8415656475453090   0.8085742077768074   0.6250000000000000
0.1342828814975405   0.1012914579325113   0.6250000000000000
```

## 1.3 C_{36}N_6

```
C36N6
   1.00000000000000
    18.6642509474898191    0.0000000000000000    0.0000000000000000
     9.3321254737449113   16.1637154671572389    0.0000000000000000
     0.0000000000000000    0.0000000000000000   16.0000000000000000
   C    N
    36    6
Direct
  0.0803805175383719   0.0134375642230721   0.6250000000000000
  0.1562770497904583   0.0123380302508641   0.6250000000000000
  0.2219428887891339   0.0127792064288285   0.6250000000000000
  0.2938493333534483   0.0127792064288285   0.6250000000000000
  0.3599563485301026   0.0123380302508641   0.6250000000000000
  0.4347533468099951   0.0134375642230721   0.6250000000000000
  0.5773223597886915   0.0195538797224047   0.6250000000000000
  0.6521193580685840   0.0206534136946198   0.6250000000000000
  0.7182263732452382   0.0202122375166483   0.6250000000000000
  0.7901328178095526   0.0202122375166483   0.6250000000000000
  0.8557986568082282   0.0206534136946198   0.6250000000000000
  0.9316951890603146   0.0195538797224047   0.6250000000000000
  0.0029796945363684   0.0908383882383877   0.6250000000000000
  0.0018801624965169   0.1667349166257611   0.6250000000000000
  0.0023213376611763   0.2324007576510538   0.6250000000000000
  0.0023213366478642   0.3043072042419853   0.6250000000000000
  0.0018801624022563   0.3704142155539266   0.6250000000000000
  0.0029796953611522   0.4452112158604362   0.6250000000000000
  0.4347533457966901   0.0908383882383877   0.6250000000000000
  0.3599563494491470   0.1667349166257611   0.6250000000000000
  0.2938493332591877   0.2324007576510538   0.6250000000000000
  0.2219428876815683   0.3043072042419853   0.6250000000000000
  0.1562770506152421   0.3704142155539266   0.6250000000000000
  0.0803805173498507   0.4452112158604362   0.6250000000000000
  0.0090960117795363   0.5877802270010367   0.6250000000000000
  0.0101955476841008   0.6625772214162233   0.6250000000000000
  0.0097543587101642   0.7286842621848079   0.6250000000000000
  0.0097543753708494   0.8005906734277659   0.6250000000000000
  0.0101955387528486   0.8662565380183719   0.6250000000000000
  0.0090960214413117   0.9421530369491019   0.6250000000000000
  0.9316951897908449   0.5877802270010367   0.6250000000000000
  0.8557986594711079   0.6625772214162233   0.6250000000000000
  0.7901328076764742   0.7286842621848079   0.6250000000000000
  0.7182263797728309   0.8005906734277659   0.6250000000000000
  0.6521193518002115   0.8662565380183719   0.6250000000000000
  0.5773223701809972   0.9421530369491019   0.6250000000000000
  0.5060378522860347   0.0923432172640238   0.6250000000000000
  0.0818853569563137   0.9406482079234658   0.6250000000000000
  0.9301903590213735   0.0923432172640238   0.6250000000000000
  0.5060378636916454   0.9406482079234658   0.6250000000000000
  0.9301903588328523   0.5164957243764050   0.6250000000000000
  0.0818853453621747   0.5164957243764050   0.6250000000000000
```

## 2. HSE06 results

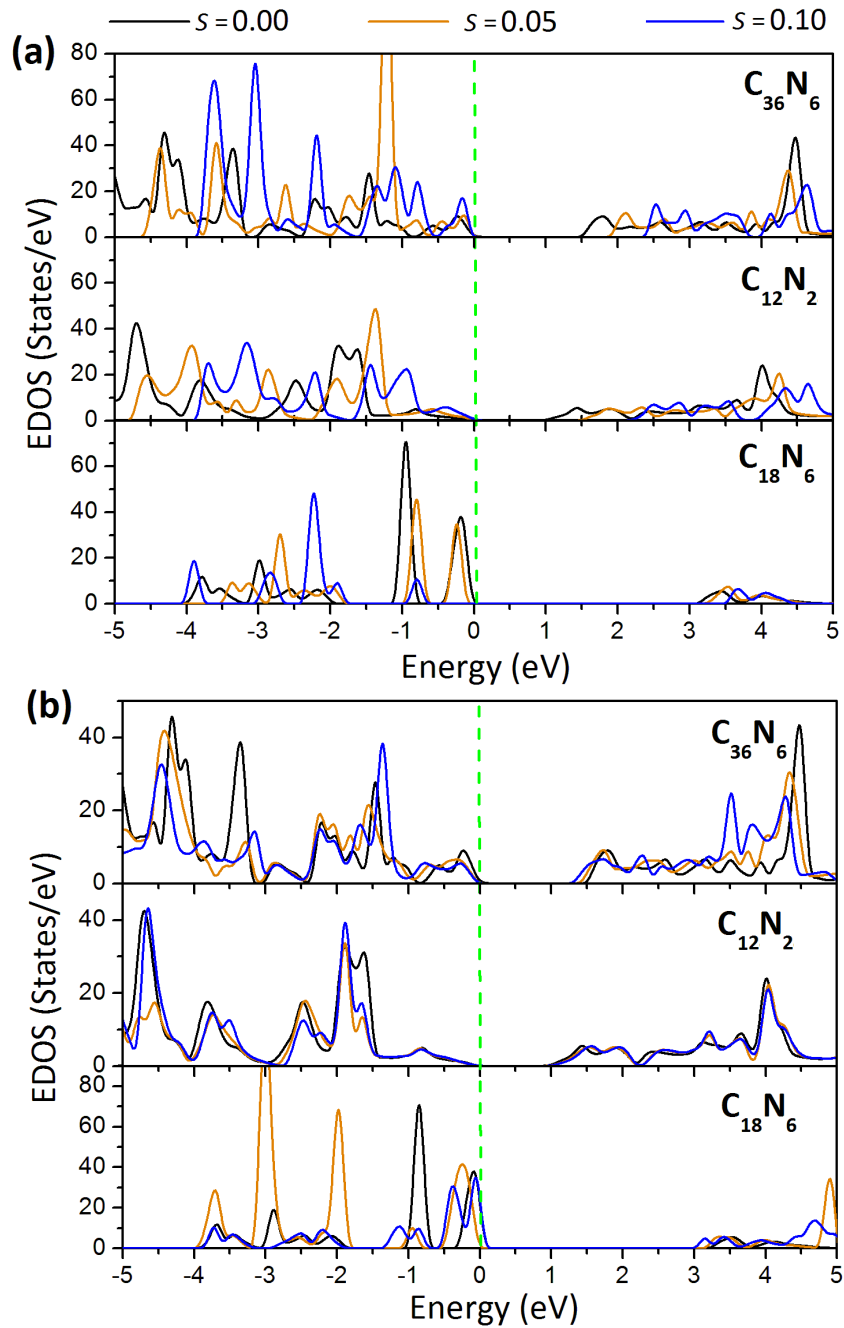

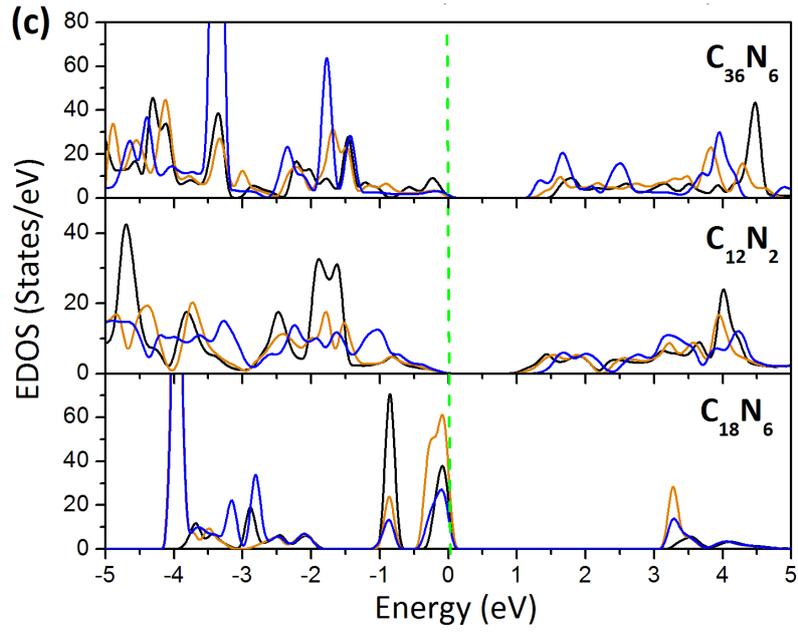

Fig. S1. The EDOS of single-layer $C_{18}B_6$, $C_{12}N_2$, $C_{36}N_6$ predicted by the HSE06 functional for different strain levels (*s*) and for the samples under the (a) biaxial loading and uniaxial loading along the (b) armchair and (c) zigzag directions. The Fermi energy is aligned to zero.